\documentstyle[prc,preprint,epsfig,aps]{revtex}
\newcommand{\ga}{\alpha}
\newcommand{\gb}{\beta}
\newcommand{\gc}{\gamma}
\newcommand{\gd}{\delta}
\newcommand{\gk}{\kappa}
\newcommand{\gve}{\varepsilon}

\newcommand{\gS}{\Sigma}

\begin{document}

\draft

\title{Deuteron formation in nuclear matter}

\author{C. Kuhrts, M. Beyer\footnote{corresponding author: FB Physik,
 Universit\"at Rostock, Universit\"atsplatz 3, 
18051 Rostock, Germany, email:beyer@darss.mpg.uni-rostock.de, Phone:
[+49] 381/4982854, Fax:[+49] 381/4982857 }, and G. R\"opke
}
\address{
FB Physik,
 Universit\"at Rostock, Universit\"atsplatz 3, 
18051 Rostock, Germany}

\vspace{10mm}

\maketitle

\begin{abstract}
  We investigate deuteron formation in nuclear matter at finite
  temperatures within a systematic quantum statistical approach.  We
  consider formation through three-body collisions relevant already at
  rather moderate densities because of the strong correlations. The
  three-body in-medium reaction rates driven by the break-up cross
  section are calculated using exact three-body equations
  (Alt-Grassberger-Sandhas type) that have been suitably modified to
  consistently include the energy shift and the Pauli blocking.
  Important quantities are the lifetime of deuteron fluctuations and
  the chemical relaxation time.  We find that the respective times
  differ substantially while using in-medium or isolated cross
  sections. We expect implications for the description of heavy ion
  collisions in particular for the formation of light charged
  particles at low to intermediate energies.
\end{abstract}

\vspace{5mm}

\section{Introduction}

The formation of deuterons is the simplest multichannel reaction
process in a heavy ion collision. Within a systematic quantum
statistical approach the formation is driven by the Boltzmann
collision integral that in turn requires the proper break-up cross
sections. Besides electromagnetic disintegration/formation of the
deuteron ($np\rightleftharpoons \gamma d$) that is a generic two-body
process, the treatment of three-body collision ($NNN\rightleftharpoons
Nd$) is more involved, and requires a solution of the three-body
dynamics in medium. Collision rates are a central input into modern
microscopic simulations of heavy ion collisions such as the
Boltzmann-Uehling-Uhlenbeck (BUU)~\cite{sto86,dan91,wolter} or the
quantum molecular dynamics approach (QMD)~\cite{QMD,QMD2}. For the
latest developments and recent applications to multifragmentation in
heavy ion collisions see Ref.~\cite{fel99}.  Danielewicz and Bertsch
~\cite{dan91} started the numerical solution of coupled BUU equations
including the deuteron formation. Later this model has been enlarged
to include the production of three-particle clusters ~\cite{Dan92}.

So far these approaches are fed with experimental cross sections, e.g.
$Nd\rightarrow NNN$ for the deuteron break-up, and the medium effect
is taken into account by the Pauli-blocking of final states. However,
in principle, as shown earlier those cross sections are changed
substantially, when medium effects are consistently
included~\cite{bey96,bey97}.  Besides including these medium dependent
cross sections in the respective simulations that is currently on its
way, it is an interesting and important question to analyze the change
of time scales induced by medium dependent reaction cross sections.

The correlated many-particle system is treated within the cluster mean
field approach~\cite{chf} that goes beyond the quasiparticle picture
and in turn has been applied to a wide variety of many-particle
physics problems. This approach is intimately related to the Dyson
Equation Approach to Many-Body Green functions~\cite{scrpa,sch73}.
Both approaches lead to a systematic decoupling of the equation
hierarchy and thus to effective few-body equations. For the
two-particle case these are known as Feynman-Galitskii or
Bethe-Goldstone equations. Recently, we have derived a corresponding
equation for the three-body system for finite
temperatures~\cite{bey96,habil} based on the formulation of Alt,
Grassberger, and Sandhas (AGS)~\cite{AGS}. For {\em zero} temperatures
three-body in-medium equations have been given in Ref.~\cite{sch73}
for the related 2p-1h problem.

The respective one-, two-, and three-body equations are derived in an
independent particle basis. That means we do not consider correlations
in the surrounding matter that would eventually lead to a
self-consistent treatment. However, such a treatment is technically
involved and only very recently progress has been achieved towards
this direction by including {\em two}-body correlations into the
one-body spectral function~\cite{boz99}. Since presently our aim is
different, i.e. to extract the formation and equilibration time
scales, and the densities considered are rather low compared to the
ones that require a self-consistent treatment, and since many technical
problems still need to be tackled we presently do not consider a such
an approach for the {\em three}-body case.

The life time and the relaxation time are determined by linearizing
the Boltzmann equation or the corresponding rate equations. Therefore
the respective integrals can be evaluated in thermal and chemical
equilibrium. To this end we assume only small fluctuations, i.e.
linear response of the system. This will be explained in the following
section. The necessary three-body reaction input on the basis of
Faddeev type equations will be given in Sect.~\ref{sec:green}. The
derivation of these equations is based on the Green function approach,
e.g. ~\cite{fet71}. In Sec.~\ref{sec:result} we will present numerical
results and in the last section we summarize and give our conclusion.

\section{Reactions}

The quantity of interest in the quantum statistical approach is the
generalized quantum Boltzmann equation for the nucleon $f_N(p,r)$,
deuteron $f_d(P,R)$, etc. Wigner distributions. For the time being we
assume symmetric nuclear matter. These distributions are driven by a
coupled system of Boltzmann equations, see, e.g. Ref.~\cite{dan91},
\begin{eqnarray}
\partial_t f_N+{\partial}_{p} U\cdot
{\partial}_{r} f_N-{\partial}_{ r} U\cdot
{\partial}_{ p} f_N&=&
{\cal K}^{\rm in}_N[f_N,f_d]\,(1-f_N)
-{\cal K}^{\rm out}_N[f_N,f_d]\, f_N,
\nonumber\\
\partial_t f_d+{\partial}_{P} U\cdot
{\partial}_{R} f_d-{\partial}_{R} U\cdot
{\partial}_{P} f_d&=&
{\cal K}^{\rm in}_d[f_N,f_d]\,(1+f_d)
-{\cal K}^{\rm out}_d[f_N,f_d]\,f_d,
\label{eqn:Boltz}
\end{eqnarray}
where $U$ denotes the total mean field potential. The coupling
between the different species is through the collision integrals
${\cal K}[f_N,f_d]$ that are merely shown for the deuteron case in the
following two equations relevant in the present context,
\begin{eqnarray}
{\cal K}^{\rm in}_d(P,t)&=&
\int d^3k\int d^3k_1d^3k_2\; |\langle kP|U|k_1k_2\rangle|^2_
{dd\rightarrow dd}\;
f_d(k_1,t)f_d(k_2,t) \bar f_d(k,t)\nonumber\\
&&+\int d^3k\int d^3k_1d^3k_2\;
|\langle kP|U|k_1k_2\rangle|^2_{Nd\rightarrow Nd}\;
f_N(k_1,t)f_d(k_2,t) \bar f_N(k,t)\nonumber\\
&&+\int d^3k\int d^3k_1d^3k_2d^3k_3\;
|\langle kP|U_0|k_1k_2k_3\rangle|^2_{pnN\rightarrow dN}\nonumber\\&&
\qquad\times
f_N(k_1,t)f_N(k_2,t)f_N(k_3,t)\bar f_N(k,t)\nonumber\\&&
+\dots\label{eqn:react},
\end{eqnarray}
where we used the abbreviations $\bar f_N=(1-f_N)$ and
$\bar f_d=(1+f_d)$,  
the ellipsis denote further possible contributions involving four and
more body collisions (e.g. $tp\rightleftharpoons dd$, $hp\rightleftharpoons
dd$) or processes like $np\rightleftharpoons \gamma d$, and 
${\cal K}^{\rm out}_d(P,t)$ is given by
\begin{eqnarray}
{\cal K}^{\rm out}_d(P,t)&=&
\int d^3k\int d^3k_1d^3k_2d^3k_3\;
|\langle k_1k_2k_3|U_0|kP\rangle|^2_{dN\rightarrow pnN}\nonumber\\&&
\qquad\times
\bar f_N(k_1,t)\bar f_N(k_2,t)\bar f_N(k_3,t)f_N(k,t)\nonumber\\&&
+\dots\label{eqn:react2}
\end{eqnarray}

The quantity $U_0$ appearing in Eqs.~(\ref{eqn:react}) and
(\ref{eqn:react2}) is the in-medium break-up transition operator for
the $Nd\rightarrow NNN$ reaction and is calculated using the AGS equation
as given in the next section.  For the isolated three-body problem
$U_0$ determines the isolated break-up cross section $\sigma_{\rm bu}^0$
via~\cite{AGS,glo88} 
\begin{eqnarray}
\sigma_{\rm bu}^0 (E)&=& \frac{1}{|v_d-v_N|} \frac{1}{3!}
\int d^3k_1d^3k_2d^3k_3\;
 |\langle kP|U_0|k_1k_2k_3\rangle|^2\nonumber\\
&&\qquad\qquad\times2\pi\delta(E'-E)\;(2\pi)^3\delta^{(3)}(k_1+k_2+k_3),
\label{eqn:sig0}
\end{eqnarray}
where $|v_d-v_N|$ denotes the relative velocity of the incoming
nucleon and deuteron.

So far the strategy has been to implement the {\em experimental} cross
section into Eqs.~(\ref{eqn:react}) and (\ref{eqn:react2}).  This has
then been solved, for a specific heavy ion collision~\cite{dan91}.
Using experimental cross sections respectively isolated cross sections
may not be sufficient in particular in the lower energy regime. The
cross section itself depends on the medium, e.g. blocking of internal
lines or self-energy corrections of the respective three-body Green
functions. This has been shown in Refs.~\cite{bey96,bey97}.  While the
inclusion of the medium dependent deuteron break-up cross sections in
a coupled BUU simulation of a heavy ion collision is a congruous
application, it is instructive to study the influence of medium
dependent cross sections in a relaxation time approximation, in
addition. This would enable us to already arrive at conclusions on the
possible effect for typical conditions at heavy ion collisions.  After
linearizing the Boltzmann equation it is possible to define a break-up
time for small fluctuations of the deuteron distributions. To do so,
we consider only the term given in Eq.~(\ref{eqn:react2}) in the
collision integral.  For small fluctuations $\gd f_d(t)=f_d(t)-f_d^0$
from the equilibrium distribution $f_d^0$ linear response leads
to~\cite{bey97}
\begin{equation}
\frac{\partial}{\partial \,t} \gd f_d(P,t) = -\frac{1}{\tau_{\rm bu}(P,n_N)}
\gd f_d(P,t),
\end{equation}
where the ``life time'' of deuteron fluctuations that depends on the
deuteron momentum $P$ and the nuclear density $n_N$ has been
introduced earlier~\cite{bey97},
\begin{eqnarray}
\tau^{-1}_{\rm bu} (P,n_N)
&=& \frac{4}{3!} \int dk_Nd^3k_1 d^3k_2 d^3k_3\;
\left|\langle kP|U_0|k_1k_2k_3\rangle\right|^2\nonumber\\
&&\; \times\bar f^0_N(k_1)\bar f^0_N(k_2)\bar f^0_N(k_3) f^0_N(k_N)
\;2\pi\gd(E - E_0),
\label{eqn:lifetime}
\end{eqnarray}
which can be related to the break-up cross section using
Eq.~(\ref{eqn:sig0}).
 
Chemical equilibration results from break-up and formation reactions, 
according to the terms that contain the transition operator $U_0$ 
in Eq.~(\ref{eqn:react}) and (\ref{eqn:react2}). The change of the total
deuteron density is given by the  integration of the deuteron momentum in
Eq.~(\ref{eqn:Boltz}). For a homogeneous system we can neglect the
drift terms and only the reaction terms will change the total density,
therefore we obtain the following rate equation 
\begin{eqnarray}
\frac{d}{dt}n_d(t)=-\alpha (t) n_N(t) n_d(t)+\beta (t) n_N^3(t),
\end{eqnarray}
where the introduced rate coefficient for disintegration $\alpha$ is
given by
\begin{eqnarray}
\alpha(t)&&=
\int d^3Pd^3k\int d^3k_1d^3k_2d^3k_3\;
|\langle k_1k_2k_3|U_0|kP\rangle|^2_{dN\rightarrow pnN}\nonumber\\&&
\qquad\times
\bar f_N(k_1,t)\bar f_N(k_2,t)\bar f_N(k_3,t))\frac{f_N(k,t)}{n_N(t)}
\frac{f_d(P,t)}{n_d(t)},\label{eqn:rate}
\end{eqnarray}
and an analogous expression holds for the formation rate $\beta$. To
obtain a time scale for the formation of chemical equilibrium, we
consider a near-equilibrium situation.
  
In chemical equilibrium both rate coefficients are constant and related by
detailed balance (time reversal invariance) resulting in
\begin{equation}
\beta=\frac{\alpha n_d^0}{(n_N^0)^2}.
\end{equation}
Here we denote the equilibrium distribution functions by $n_N^0$ and
$n_d^0$.  Assuming now small density fluctuations, viz.
$n_N=n_N^0+\delta n_N$ and $n_d=n_d^0-\delta n_N/2$, and after
insertion into Eq.~(\ref{eqn:Boltz}) a chemical relaxation time may be
defined by
\begin{equation}
\frac{d}{dt}\, \gd n_d(t) = -\frac{1}{\tau_{\rm rel}(n_N)}\,
\gd n_d(t).
\end{equation}
Using the definition of the rate coefficient $\alpha$ given in
Eq.~(\ref{eqn:rate}) leads to following expression for the chemical 
relaxation time
\begin{equation}
\tau^{-1}_{\rm rel}(n_N^0) = 
\int d^3k\;d^3P\;f^0_N(k)f^0_d(P) |v_d-v_N|\; \sigma_{\rm bu}(k,P)
\;\frac{n^0_N+4n^0_d}{n^0_Nn^0_d},
\label{eqn:chem}
\end{equation}
The deuteron break-up cross section $\sigma_{\rm bu}$ is defined as in
Eq.~(\ref{eqn:sig0}) but may now also include medium modifications of the
transition operator $U_0$ as will be explained in the next section.
The deuteron equilibrium density $n^0_d$ is related to the nucleon
density via $\mu_d=2\mu_N$ for the chemical potentials.

\section{Finite temperature three-body equations}
\label{sec:green}

The basis to derive effective in-medium few-body Green functions for
finite temperatures is provided by the cluster mean field
approximation. This approximation allows us to truncate the hierarchy
of Green function equations in a controlled way. A general
introduction may be found in Ref~\cite{chf}. This approach is related
to the Dyson Equation Approach~\cite{scrpa,sch73} extended here to finite
temperatures~\cite{habil,sch97}. It has been proven useful in the
context of strongly coupled systems provided, e.g., in nuclear
physics~\cite{sch73}.  A calculable form is achieved by introducing
ladder approximation. We consider generic elementary two-body
interactions $V_2$ only. The resulting equations for the decoupled
one-, two- and three-body Green functions at finite temperatures
(utilizing the Matsubara technique to treat finite
temperatures~\cite{fet71}) will be given in the following. The {\em
  one-particle} Green function reads
\begin{equation}
G_1(z) = R_1^{(0)}(z) = \left(z - \varepsilon_1\right)^{-1}.
\label{eqn:G1}
\end{equation}
In mean field approximation the single quasiparticle
energy $\varepsilon_1$ is given by
\begin{eqnarray}
\varepsilon_1 &= &\frac{k^2_1}{2m_1} + \Sigma^{HF}(1),\nonumber \\
\Sigma^{HF}(1) &=& \sum_{2}
\left[ V_2(12,12)-  V_2(12,21) \right]\,  f_2,
\label{eqn:selfHF}
\end{eqnarray}
and $f_2=f^0_N(\epsilon_2)$ the Fermi function\footnote{Here and in the
  rest of the paper we use equilibrium distribution functions only,
  and therefore we use $f^0_N\rightarrow f$ for notational simplicity.}.
This is given by
\begin{equation}
f(\omega) = \frac{1}{e^{\beta(\omega-\mu)}+1},
\end{equation}
where $\beta$ denotes the inverse temperature and $\mu$ the chemical
potential. The equation for the {\em two-particle} Green function
$G_2(z)$ reads
\begin{equation}
G_2(z) = N_2R_2^{(0)}(z) + R_2^{(0)}(z)N_2V_2 G_2(z),
\label{eqn:G2}
\end{equation}
Note that we use a matrix notation for the Fock space indices
($1,2,3,\dots$) for convenience. The matrix for the 
two-body resolvent $R_2^{(0)}(z)$ is then given by
\begin{equation}
 R_2^{(0)}(12,1'2';z) = \frac{\delta_{11'}\delta_{22'}}
{z - \varepsilon_1 - \varepsilon_2}
\end{equation}
and the Pauli blocking factor $N_2$ by
\begin{eqnarray}
N_2(12,1'2') &= &\delta_{11'}\delta_{22'}
(\bar f_1 \bar f_2 - f_1 f_2)\nonumber\\ 
&=&\delta_{11'}\delta_{22'}
(1 - f_1 -  f_2).
\end{eqnarray}
We use the abbreviation $\bar f = 1-f$.  The respective equation for
the {\em three-particle} Green function relevant to describe
three-body correlations in a medium is given by
\begin{equation}
G_3(z) = N_3R_3^{(0)}(z) + R_3^{(0)}(z) [N_2V_2] G_3(z),
\label{eqn:G3}
\end{equation}
where we have introduced the notation
\begin{eqnarray}
  [N_2V_2](123,1'2'3') &=&\sum_{\gamma=1}^3 
(N_2V_2)^{(\gamma)}(123,1'2'3'),
\label{eqn:Vchn}\\
(N_2V_2)^{(3)}(123,1'2'3')&=& (1-f_1-f_2) V_2(12,1'2')\delta_{33'}.
\label{eqn:V3}
\end{eqnarray}
The last equation is given for $\gamma=3$. The channel notation is
convenient to treat systems with more than two particles~\cite{glo88}.
In the three-particle system usually the index of the spectator
particle is used to characterize the channel. The Pauli factors $N_3$
and the resolvents $R_3^{(0)}$ read respectively
\begin{eqnarray}
N_3(123,1'2'3') &= &\delta_{11'}\delta_{22'}\delta_{33'}
(\bar f_1 \bar f_2 \bar f_3 + f_1 f_2  f_3)\label{eqn:FPauli}\\
 R_3^{(0)}(123,1'2'3';z) &= &\frac{\delta_{11'}\delta_{22'}\delta_{33'}}
{z - \varepsilon_1 - \varepsilon_2 - \varepsilon_3}.
\end{eqnarray}
Note that $[N_2V_2]\neq [N_2V_2]^\dagger$ and
$N_3R_3^{(0)}=R_3^{(0)}N_3$.  Although we assume a fermionic system
the proper symmetrization of the equations is treated separately.

If the correlated pair, e.g. $'12'$, and the spectator particle, e.g. $'3'$,
are uncorrelated in the channel $(3)$ we may define a channel Green
function $G_3^{(3)}(z)$. We generalize to the channel $(\gamma)$. The 
channel Green function is then defined by
\begin{equation}
G_3^{(\gamma)}(z) = \frac{1}{-i\beta} \sum_\lambda\;
iG_2(\omega_\lambda)\;G_1(z-\omega_\lambda). 
\label{eqn:Gchanneldef}
\end{equation}
The summation is done over the bosonic Matsubara frequencies
$\omega_\lambda$, $\lambda$ even, $\omega_\lambda=\pi\lambda/(-i\beta)
+2\mu$.  The equation for the channel Green function is derived in the
same way as for the total three-particle Green function given in
Eqs.~(\ref{eqn:G3}). The result is (no summation of $\gamma$)
\begin{equation}
G_3^{(\gamma)}(z) = N_3R_3^{(0)}(z) + R_3^{(0)}(z)
  (N_2V_2)^{(\gamma)} G_3^{(\gamma)}(z).
\label{eqn:Gchannel}
\end{equation}
We are now ready to define the proper transition operator $U_{\ga\gb}$,
\begin{equation}
G_3(z)=\gd_{\ga\gb} G_3^{(\ga)}(z)+ 
G_3^{(\ga)}(z)U_{\ga\gb}(z)G_3^{(\gb)}(z),
\end{equation}
which in the zero density limit coincides with the usual
definition of the transition operator with the correct reduction
formula to calculate cross sections~\cite{san72}. 
After proper three-body algebra we arrive at the following equation
for the transition operator in medium, viz.
\begin{equation}
N_3U_{\ga\gb}(z) = (1-\gd_{\ga\gb}) [R_3^{(0)}(z)]^{-1}+ 
\sum_{\gc\neq \ga} (N_2V_2)^{(\gc)} G_3^{(\gc)}(z)U_{\gc\gb}(z).
\end{equation}
For $N_3\neq 0$ the equation may be multiplied by $N_3^{-1}$ from the
left. We now connect this equation to the two-body $t$ matrix.  To
this end we define the following transition channel operator
$T^{(\gc)}_3(z)$,
\begin{equation}
G_3^{(\gc)}(z)= N_3R_3^{(0)}(z)+ 
R_3^{(0)}(z) N_2^{(\gc)} T^{(\gc)}_3(z) N_3 R_3^{(0)}(z).
\label{eqn:Tdef}
\end{equation}
The definition has been chosen so that is leads to the standard
Bethe-Goldstone equation in the two-nucleon subchannel,
$T_3^{(3)}(z)=T_2(\tilde z)\gd_{33'}$, where $\tilde z$ is
the frequency of the two-body subsystem embedded in the three-body
system. Inserting eq.~(\ref{eqn:Tdef}) into eq.~(\ref{eqn:Gchannel})
leads to the subchannel Bethe-Goldstone equation
\begin{equation}
  T^{(\gc)}_3(z) =   V_2^{(\gc)} + 
(V_2N_2)^{(\gc)} R_3^{(0)}(z)   T_3^{(\gc)}(z).
\label{eqn:Tchannel}
\end{equation}
With the help of this equation it is possible to connect the AGS type
equation directly to the two-body sub $t$ matrix
\begin{equation}
N_3U_{\ga\gb}(z) = (1-\gd_{\ga\gb}) [R_3^{(0)}(z)]^{-1}+ 
\sum_{\gc\neq \ga}   N_2^{(\gc)} T_3^{(\gc)}(z) R_3^{(0)}(z) N_3
U_{\gc\gb}(z). 
\label{eqn:AGS}
\end{equation}
Here we have written the Pauli factors occurring  due to the
surrounding matter explicitly. Note, that through eq.~(\ref{eqn:Tchannel})
$T_3^{(\gc)}(z)$ is also medium dependent.  

Since $N_3>0$ we may multiply Eq.~(\ref{eqn:AGS}) by $N_3^{-1/2}$ form 
the left and by $N_3^{1/2}$ from the right, and introduce
$U^*_{\gc\gb}=N^{1/2}_3U_{\gc\gb}N^{1/2}_3$. Then
\begin{equation}
U^{*}_{\ga\gb}(z)
= (1-\gd_{\ga\gb}) [R_3^{(0)}(z)]^{-1}+ 
\sum_{\gc\neq \ga}  N_3^{-1/2} N_2^{(\gc)} T_3^{(\gc)}(z) 
N_3^{1/2} R_3^{(0)}(z) U^*_{\gc\gb}(z).
\label{eqn:AGSlow}
\end{equation}
We may now introduce the transition channel operator $T_3^{*(\gc)} =
N_3^{-1/2} N_2^{(\gc)} T_3^{(\gc)} N_3^{1/2}$ so that the equation
looks formally as for the isolated case. The respective channel $t$
matrix equation then is
\begin{equation}
T_3^{*(\gc)}= V_3^{*(\gc)}+V_3^{*(\gc)}R_3^{(0)}T_3^{*(\gc)}.
\label{eqn:Tchstar}
\end{equation}
The explicit form of the effective potential arising in this equation
reads (e.g. in the (3) channel) 
\begin{eqnarray}
V_3^{*(3)}(123,1'2'3')&=&  N_3^{-1/2}(123) N_2^{(3)}(123) 
V_3^{(3)}(123,1'2'3')N_3^{1/2}(1'2'3')
\nonumber\\
&=&(1-f_1-f_2)^{1/2}(1-f_3+g(\gve_1+\gve_2))^{-1/2} \nonumber\\
&&\times V_2(12,1'2')\gd_{33'}
(1-f_3+g(\gve_{1'}+\gve_{2'}))^{1/2}(1-f_{1'}-f_{2'})^{1/2}\\
&\simeq&(1-f_1-f_2)^{1/2} V_2(12,1'2')(1-f_{1'}-f_{2'})^{1/2}.
\label{eqn:Vapprox}
\end{eqnarray}
The last equation holds for $f^2\ll f$ only and
$g(\omega)=[e^{\beta(\omega-2\mu)}-1]^{-1}$. Utilizing this
approximation eq.~(\ref{eqn:AGSlow}) has been solved numerically using
a separable potential for the strong nucleon-nucleon
interaction~\cite{bey96}.

\section{Results}
\label{sec:result}

Since the Green functions have been evaluated in an independent
particle basis the one-, two-, and three-particle Green functions are
decoupled in hierarchy, as given in Eqs.~(\ref{eqn:G1}),
(\ref{eqn:G2}), (\ref{eqn:G3}). To solve the in-medium problem up to
three-particle clusters the one-, two- and three-particle problems are
consistently solved. This leads to the single particle self-energy
shift Eq.~(\ref{eqn:selfHF}), the two-body input including the proper
Pauli blocking Eq.~(\ref{eqn:Tchstar}), and eventually to the
three-body scattering state.

For technical reasons we have used some reasonable approximations
valid at smaller effective densities. The
nucleon self-energy has been calculated via Eq.~(\ref{eqn:selfHF}),
but in the three-body code we have approximated the self-energy by
effective masses, i.e.
\begin{equation}
\gve_1=\frac{k_1^2}{2m}+\gS^{HF}(k_1)
\simeq \frac{k_1^2}{2m^*}+\gS^{HF}(0).
\end{equation}

For simplicity and speed of the calculation, we
use angle averaged Fermi functions $\langle\langle N_2\rangle\rangle$
and $\langle\langle N_3\rangle\rangle$, e.g.
\begin{equation} 
\langle\langle N_3\rangle\rangle =
\frac{1}{(4\pi)^2}\;\int d\cos\theta_q \; d\cos\theta_p \; d\phi_q
d\phi_p\; N_3({p},{q},{P}_{\rm c.m.}), 
\end{equation}
where the angles are taken with respect to ${P}_{\rm c.m.}$ and
${p}$, ${q}$ are the standard Jacobi coordinates~\cite{glo88}.

We have solved the three-body equation neglecting terms of the order
$f^2$, using eq.~(\ref{eqn:Vapprox}) as the driving kernel in the AGS
equations, which is sufficient, since the calculation is done below
the Mott density of the deuteron.

\begin{figure}[t]
  \begin{center}
\psfig{figure=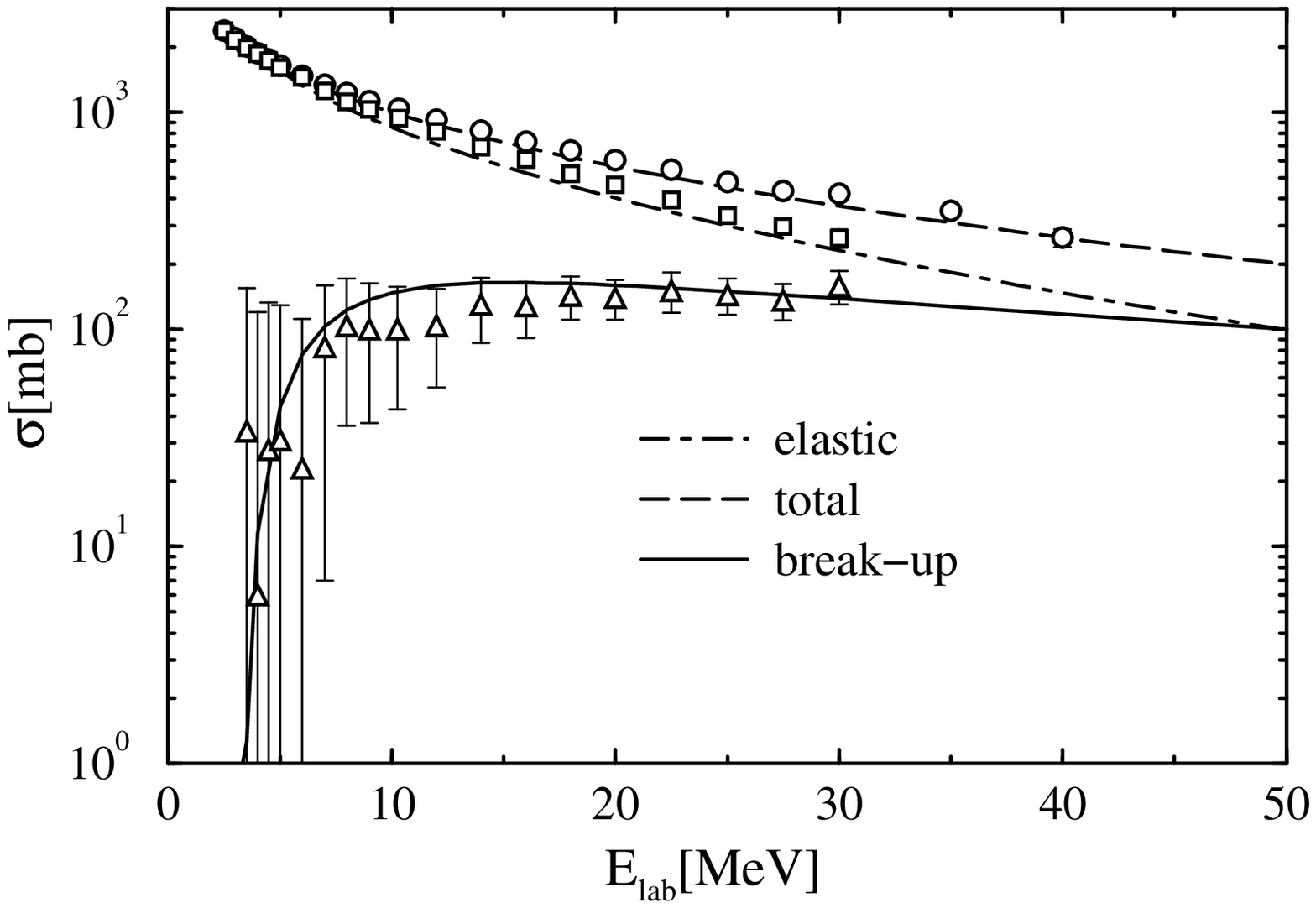,width=0.7\textwidth}
\caption{\label{fig:isocross}  A comparison of the total, elastic,  
  and break-up cross sections $nd\rightarrow nd$, $nd\rightarrow
  nnp$ with the experimental data of
  Ref.~\protect\cite{sch83}.}
  \end{center}
\end{figure}

A rank-one Yamaguchi potential \cite{yamaguchi} has been used in the
calculations of the deuteron break-up cross section that includes the
coupled $^3$S$_1$-$^3$D$_1$ and the $^1$S$_0$ channels only. The
parameters are given in Table~\ref{tab:parm}.  To get an
impression of the quality of the calculation the isolated cross
section is given in Fig.~\ref{fig:isocross} along with the
experimental data on neutron deuteron scattering~\cite{sch83}.

Unlike the isolated three-body problem the integral equation depends
on the relative momentum of the three-nucleon system with
respect to the medium, since the Fermi functions depend explicitly on
the center of mass momentum of the three-body system.  As a
consequence one has to solve the three-body problem at a finite center
of mass momentum $P_{\rm c.m.}=k_1+k_2+k_3$ in a medium that may be
considered at rest $P_{\rm med}=0$.  However, technically it is more
convenient to let the three-nucleon system rest $P_{\rm c.m.}=0$ and the
surrounding medium move with $P_{\rm med} = - (k_1+k_2+k_3)$. This
procedure results in the least change of the three-body algebra and is
possible since the dependence on $P_{\rm c.m.}-P_{\rm med}$ is only
through the Pauli blocking factors thus parametric.

\begin{figure}[t]
  \begin{center}
\psfig{figure=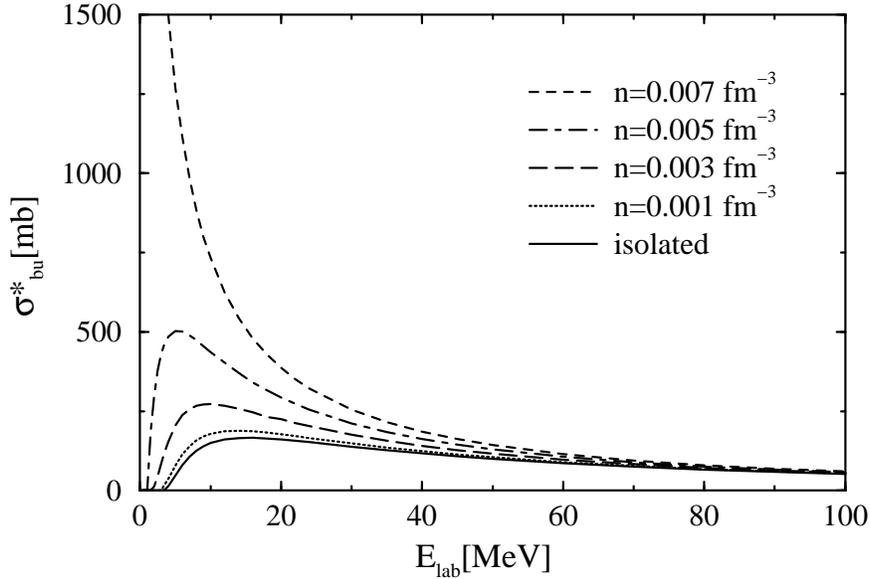,width=0.7\textwidth}
\caption{\label{fig:cross} In-medium break-up cross section at $T=10$
  MeV and $P_{\rm c.m.}=0$.
  The isolated cross section is shown as a solid line, other lines  
  show different nuclear densities as given in the legend.  }
  \end{center}
\end{figure}

\begin{figure}[t]
  \begin{center}
\psfig{figure=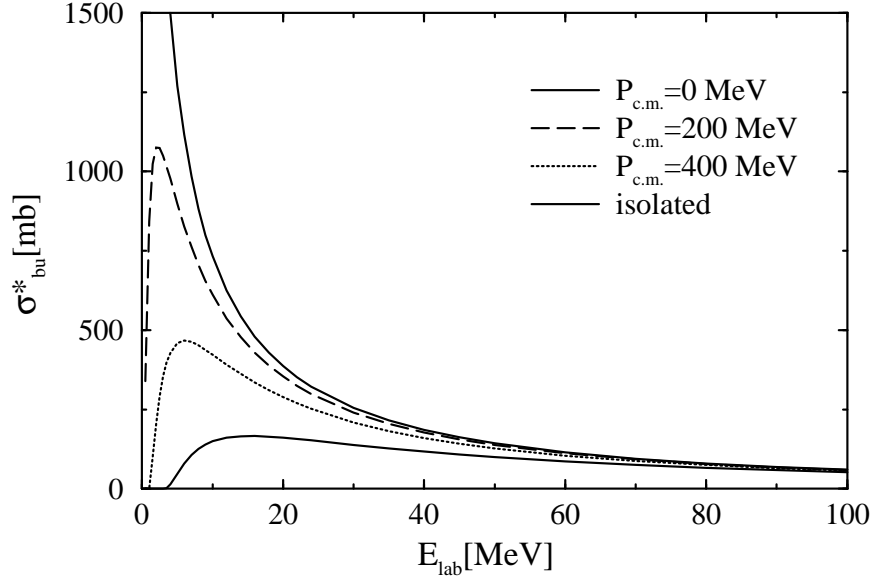,width=0.7\textwidth}
\caption{\label{fig:crossP} In-medium break-up cross section at $T=10$
  MeV and $n=0.007$ fm$^{-3}$ for different center-of-mass momenta
  $P_{\rm c.m.}$. 
 The isolated cross section $\sigma^0_{\rm bu}$ is shown as solid line.}
  \end{center}
\end{figure}

\begin{figure}[tbh]
  \begin{center}
\psfig{figure=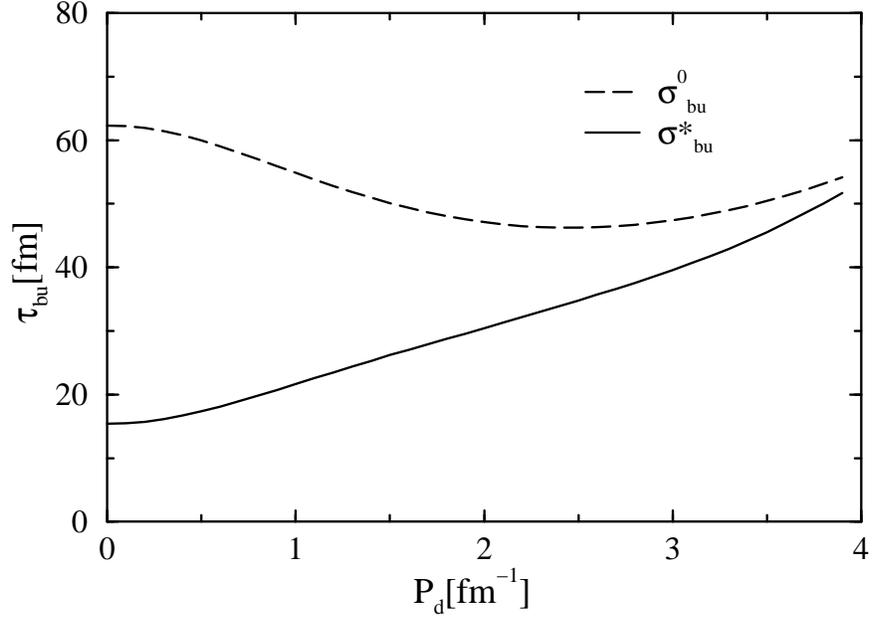,width=0.7\textwidth}
\caption{\label{fig:life} Deuteron break-up time at $T=10$ MeV and  
  nuclear density $n=0.007$ fm$^{-3}$. Solid line using the medium
  dependent cross section as given in Fig.~\protect\ref{fig:cross},
  and dashed line using the isolated cross
  section. } 
  \end{center}
\end{figure}

For small momenta $P_{\rm c.m.}$ and small relative energies of the
$dN$-system the influence of the medium is most pronounced.  In
Fig.~\ref{fig:cross} we see that the in-medium cross section is
significantly {\it enhanced} compared to the isolated on.  So it is
not a justifiable approximation to include medium effects only by the
Pauli-blocking factors of the final states (see
eq.~(\ref{eqn:react2})). That would effectively decrease the break-up
cross section and might lead to wrong conclusions.  The dominant
effect is the weakening of deuteron binding energy (Mott-effect). This
provokes also a shift of the threshold to smaller energies with
raising densities.  We observe that for higher energies the medium
dependence of the cross section becomes much weaker, which {\em a
  posteriori} justifies the use of isolated cross sections (along with
the impulse approximation) when higher energies are
considered~\cite{dan91}. The dependence of the cross section on the
center-of-mass momentum $P_{\rm c.m.}$ is shown in
Fig.~\ref{fig:crossP} for a fixed density and temperature.  With the
definition of a medium dependent cross section according to
Eq.~(\ref{eqn:sig0}) we can rewrite for the deuteron break-up time
Eq.~(\ref{eqn:lifetime})
\begin{equation}
\tau^{-1}_{\rm bu}(P) 
= 4 \int dk_N\;
|v_d-v_N|\sigma_{\rm bu}(k,P)f(k_N).
\label{eqn:lifetime2}
\end{equation}
To get an impression how much the medium dependence of the cross
section shortens the deuteron break-up time we show in
Fig.~\ref{fig:life} the result of the above Eq.~(\ref{eqn:lifetime2})
with use of the medium cross section $\sigma^{*}_{\rm bu}$ calculated
by solving Eq.~(\ref{eqn:AGSlow}) compared to the isolated
$\sigma^0_{\rm bu}$. For small deuteron momenta the break-up time is
shorter by a factor of three at $n=0.007$ fm$^{-3}$ and $T=10$ MeV..

\begin{figure}[t]
  \begin{center}
\psfig{figure=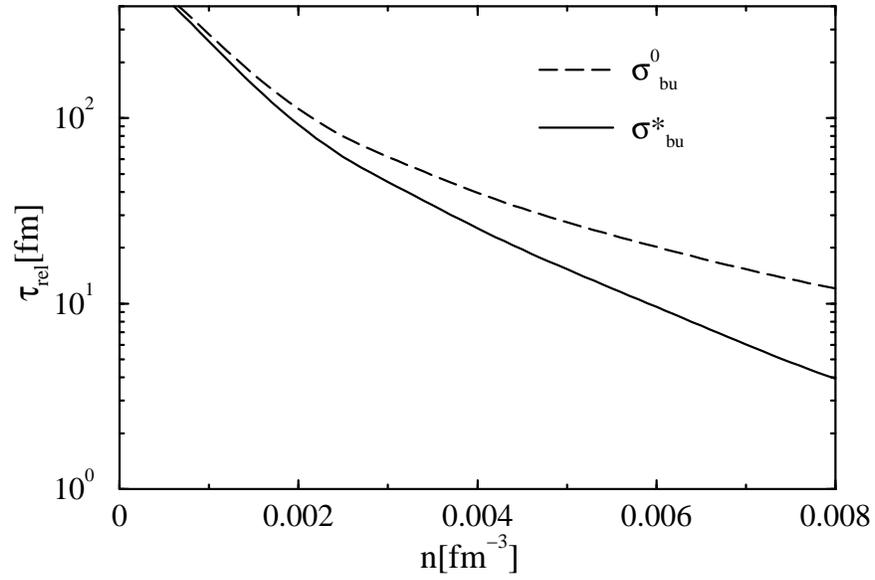,width=0.7\textwidth}
\caption{\label{fig:Trel} Relaxation time for small fluctuations of
  the deuteron density from chemical equilibrium at a temperature of
  $T=10$ MeV as a function of the nucleon density. Line coding as in
  Fig.~\protect\ref{fig:life}.}
  \end{center}
\end{figure}

We also calculate the chemical relaxation time given in
Eq.~(\ref{eqn:chem}). In  Fig.~\ref{fig:Trel} we see again that the
medium dependence of the cross section shortens the time scale
especially at higher densities.

\section{Conclusion and Outlook}

Our results show that medium dependent cross sections in the
respective collision integrals lead to {\em shorter reaction time
  scales}.  Chemical processes become faster. This also effects the
elastic rates that are related to thermal equilibration.

The basis of this result is the cluster Hartree-Fock approach that in
our approximation includes correlations up to three particles in a
consistent way. The equations driving the correlations are rigorous.
The respective one-, two- and three-body equations are solved, in
particular for the three-particle case Faddeev/AGS type equations
have been given in Ref.~\cite{bey96,bey97,habil}. The AGS approach is
particularly appealing since it allows generalizations to $n$-particle
equations in a straight forward way. Results for the three-body bound
state in medium will be published elsewhere~\cite{bey99}. As
expected from the deuteron case, the triton binding energy changes
with increasing density up to the Mott density, where $E_t=0$.

The production rates, distribution of fragments, spectra etc.  of
light charged particles in heavy ion collisions at intermediate
energies may change because of the much smaller time scales induced
through the medium dependence compared to the use of free cross
sections (respectively experimental cross sections).

\section{Acknowledgment}
We thank A. Sedrakian for a useful discussion in the early stage of
this work. This work has been supported by the Deutsche
Forschungsgemeinschaft.


\begin{table}[ht]
\caption{\label{tab:parm} Parameters for the separable
  potential used to solve the three-body AGS equation. The form
  factors are $g_{\gk\ell}(p)= 
  C_{\gk\ell} 
 p^\ell/(p^2+\gb^2_{\gk\ell})^{\ell/2+1}$. The binding energy of the
 isolated deuteron is $E_d= 2.225$ MeV and of the isolated triton
 $E_t=7.937$ MeV~{\protect\cite{bey99}}.} 
\[
\begin{array}{ccccccc} 
\hline\hline
           \gk    &C_0  &C_2      &\gb_0   &\gb_1 \\
&\multicolumn{2}{c}{\mbox{[MeV fm$^{-1}$]}}
&\multicolumn{2}{c}{\mbox{[fm$^{-1}$]} }\\
\hline
^3\mbox{S}_1- {^3\mbox{D}_1} & 8.63237& -38.8017&1.24128 &1.94773 \\
^1\mbox{S}_0       & 8.87026&         &1.17328 &       \\
\hline\hline
\end{array}
\]
\end{table}

\end{document}